# Performance Evaluation of SCM-WDM System Using Different Linecoding

**Md. Shamim Reza, Md. Maruf Hossain, Adnan Ahmed Chowdhury, S. M. Shamim Reza** and **Md. Moshiur Rahman**

**Abstract -** This paper investigates the theoretical performance analysis for a subcarrier multiplexed (SCM) wavelength division multiplexing (WDM) optical transmission system in presence of optical beat interference (OBI) which occurs during the photo detection process. We have presented a comparison for improving the performance of SCM-WDM system in presence of OBI. Non-return-to zero (NRZ), Manchester and Miller code (MC) line coding are used for performance investigation of SCM-WDM system. A suitable signal bandwidth is selected and 200 KHz is considered as channel bandwidth. Power spectrum of signal and cross component for those line coding are analyzed. Comparison results are evaluated in terms of signal to OBI ratio for the three linecoding schemes which is called signal to interference ratio (SIR). It is found that there is a significant increase in the SIR by employing Miller code compared to NRZ and Manchester for the same data rate. For example, for a number of subcarriers of 10, the achievable SIR is about -24 dB for Miller coded system compared to -46 dB for NRZ coded system and -49 dB for Manchester coded system. The results are found to be satisfactorily agreed with the expected results.

——————————— ◆ ———————————

## 1 INTRODUCTION

IN telecommunications and computer networks, multiplexing is a process where multiple analog message signals or digital data streams are combined into one signal over a shared medium. The aim is to share an expensive resource. For example, in telecommunications, several phone calls may be transferred using one wire. It originated in telegraphy, and is now widely applied in communications. The multiplexed signal is transmitted over a communication channel, which may be a physical transmission medium. The multiplexing divides the capacity of the low-level communication channel into several higher-level logical channels, one for each message signal or data stream to be transferred. A reverse process, known as demultiplexing, can extract the original channels on the receiver side. There are three fundamental types of multiplexing, all of which have significant variations. These main types are Space-Division Multiplexing (SDM), Frequency-division multiplexing (FDM) and Time-Division Multiplexing (TDM).

In order to use the optical bandwidth provided by optical fibers more efficiently, new transmission technologies have been developed in recent years, such as time division multiplexing (TDM), wavelength division multiplexing (WDM), and their combinations **[1]**.

The demand for ultra-large capacity transmission systems for long-haul communications systems and future multi-media services is increasing rapidly **[2]**. Multiplexing of optical signals promises to meet this demand. Optical transmission with subcarrier multiplexing (SCM) is a scheme where multiple signals are multiplexed in the radio frequency (RF) domain and are used to modulate directly a laser diode. Because of the simple and low-cost implementation, SCM has also been proposed to transmit multichannel digital optical signals using direct detection [2], [3] for local area optical networks. Several approaches being possible depending on the application of subcarrier multiplexing (SCM) which offers great application potential [2], [3]. Performance of an optical transmission system with RF subcarrier multiplexing (SCM) is affected by the beat interference generated due to beating of the subcarrier frequency components during the photo detection process which limits the allowable subcarrier power, the optical modulation index and maximum number of subcarrier channels that can be multiplexed in an optical channel for transmission over a single mode fiber (SMF). Researchers have been carried out recently to evaluate the impact of OBI on the performance of an SCM WDM system [4], [5].

A significant advantage of SCM is that microwave devices are more mature than optical devices; the stability of a microwave oscillator and the frequency selectivity of a microwave filter are

————————————————

- *Md. Shamim Reza is with the Dept. of Electrical and Electronic Engineering, Bangladesh University of Engineering and Technology, Dhaka-1000, Bangladesh.*
- *Md. Maruf Hossain is with the Dept. of Electrical and Electronic Engineering, American International University of Bangladesh, Dhaka, Bangladesh.*
- *Adnan Ahmed Chowdhury is with the Dept. of Electrical and Electronic Engineering, International Islamic University of Chittagong, Dhaka Campus, Bangladesh.*
- *S. M. Shamim Reza is with the Dept. of Electrical and Electronic Engineering, International Islamic University of Chittagong, Dhaka Campus, Bangladesh.*
- *Md. Moshiur Rahman is with the Dept. of Electrical and Electronic Engineering, International Islamic University of Chittagong, Dhaka Campus, Bangladesh.*

much better than their optical counterparts. In addition, the low phase noise of RF oscillators makes coherent detection in the RF domain easier than optical coherent detection, and advanced modulation formats can be applied easily. A popular application of SCM technology in fiber optic systems is analog cable television (CATV) distribution [6], [7]. Apart from possibility of combining digital and analog subcarrier multiplexed signals into a composite signal, an alternative attractive strategy is the so-called hybrid SCM system which combines a baseband digital signal with a high frequency composite microwave signal **[8]**. In this case the receiver cannot be narrowband but must have a bandwidth from d.c. to beyond the highest microwave signal frequency employed.

We start by briefly describing the linecoding used in communications in section 2. In Section 3, we have presented the system block diagram of SCM-WDM system. Mathematical analysis of signal to interference ratio (SIR) is presented in section 4. Comparison results are presented in section 5 for different linecoding used in SCM-WDM system. Finally we give our conclusion in section 6.

## 2    LINE CODING

In telecommunication a line code is a code chosen for use within a communication system for transmission purpose. For digital data transport line coding is often used. Line coding consists of representing the digital signal to be transported, by an amplitude- and time-discrete signal that is optimally tuned for the specific properties of the physical channel (and of the receiving equipment). The waveform pattern of voltage or current used to represent the 1s and 0s of a digital signal on a transmission link is called line encoding. After line coding, the signal can directly be put on a transmission line, in the form of variations of the current. Binary line codes are generally preferred because of the larger bandwidth available in optical fiber communications. In addition, these codes are less susceptible to any temperature dependence of optical sources and detectors. Under these conditions two level codes are more suitable than codes which utilize an increase number of levels (multilevel codes). Nevertheless, these factors do not entirely exclude the use of multilevel codes, and it is like that ternary codes (three levels 0,1/2,1) which give increased information transmission per symbol over binary codes will be considered for some system applications. The corresponding symbol transmission rate (i.e. bit rate) for a ternary code may be reduced by a factor of 1.58 ($\log_2 3$), whilst still providing the same information transmission rate as a similar system using a binary codes. It must be noted that this gain in information capacity for a particular bit rate is obtained at the expense of the dynamic range between adjacent levels as there are three levels inserted in place of two. This is exhibited as a 3dB SNR penalty at the receiver when compared with a binary system at a given BER. Therefore ternary codes (higher multilevel codes) are not attractive for long haul systems. For the reason described above most digital optical fiber communication systems currently in use employ binary codes. In practice, binary codes are designed which insert extra symbols into the information data stream on a regular and logical basis to minimize the number of consecutive identical received symbols, and to facilitate efficient timing extraction at the receiver by producing a high density of decision level crossings. The reduction in consecutive identical symbols also helps to minimize the variation in the mean signal level which provides a reduction in the low frequency response requirement of the receiver. These shapes are transmitted signal spectrum by reducing the d.c. component. However, this factor is less important for optical fiber systems where a.c. coupling is performed with capacitors, unlike metallic cable systems where transformers are often used, and the avoidance of d.c. components is critical. A future advantage is apparent within the optical receiver with a line code which is free from long identical symbol sequences, and where the continuous presence of 0 and 1 levels aids decision level control and avoids gain instability effects.

A non-return-to-zero (NRZ) line code is a binary code in which 1's are represented by one significant condition (usually a positive voltage) and 0's are represented by some other significant condition (usually a negative voltage), with no other neutral or rest condition. For a given data signaling rate, *i.e.*, bit rate, the NRZ code requires only half the bandwidth required by the Manchester code. When used to represent data in an asynchronous communication scheme, the absence of a neutral state requires other mechanisms for data recovery, to replace methods used for error detection when using synchronization information when a separate clock signal is available. NRZ-Level itself is not a synchronous system but rather an encoding that can be used in either a synchronous or asynchronous transmission environment, that is, with or without an explicit clock signal involved. Because of this, it is not strictly necessary to discuss how the NRZ-Level encoding acts "on a clock edge" or "during a clock cycle" since all transitions happen in the given amount of time representing the actual or implied integral clock cycle. Power spectral density of NRZ is shown in fig 1.

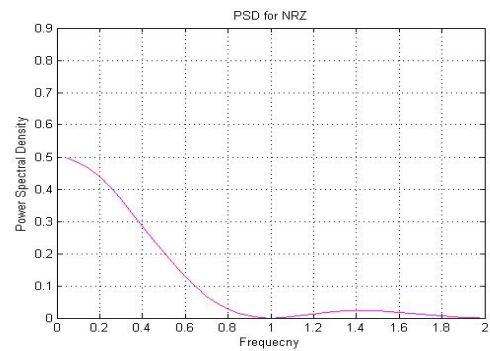

Fig. 1. PSD for NRZ.

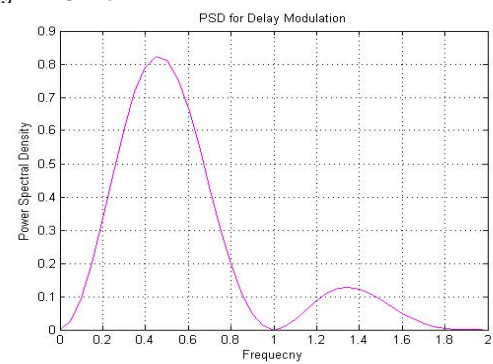

Fig. 2. PSD for Delay modulation.

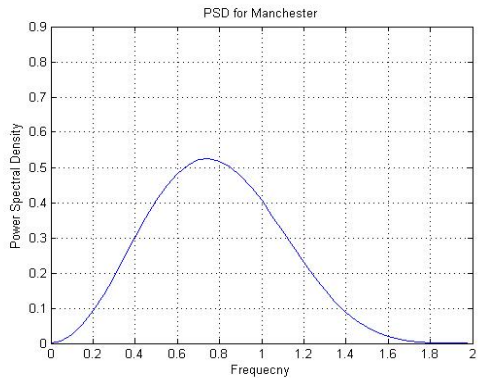
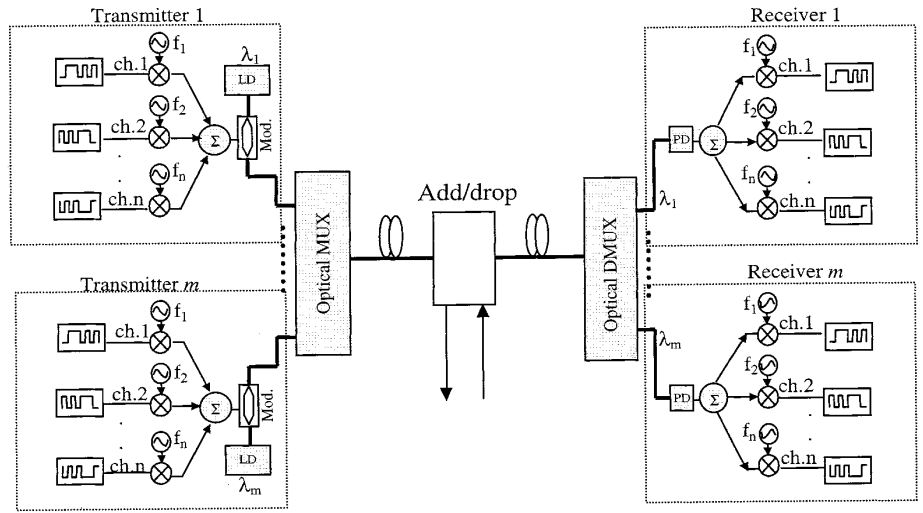

Fig. 3. PSD for Manchester Coding
Fig.4. SCM/WDM system architecture

DM encoding is also known as Miller code. The modulation has four encoder states and the source is binary. Four base band signals are available. Which one is transmitted depends both on the present digit and the previous transmitted bit. Each sate is identified by label of the previously transmitted signal. Power spectrum of delay modulation is shown in fig. 2.

In telecommunication, Manchester code (also known as Phase Encoding or PE) is a line code in which the encoding of each data bit has at least one transition and occupies the same time. It is, therefore, self-clocking, which means that a clock signal can be recovered from the encoded data. Manchester code is widely-used. There are more complex codes e.g. 8B/10B encoding which use less bandwidth to achieve the same data rate (but which may be less tolerant of frequency errors and jitter in the transmitter and receiver reference clocks). Manchester code ensures frequent line voltage transitions, directly proportional to the clock rate. This helps clock recovery. The DC component of the encoded signal is not dependent on the data and therefore carries no information, allowing the signal to be conveyed conveniently by media (e.g. Ethernet) which usually do not convey a DC component. Manchester code always has a transition at the middle of each bit period and may (depending on the information to be transmitted) have a transition at the start of the period also. The direction of the mid-bit transition indicates the data. Transitions at the period boundaries do not carry information. They exist only to place the signal in the correct state to allow the mid-bit transition. Although this allows the signal to be self-clocking, it doubles the bandwidth requirement compared to NRZ coding schemes. In the Thomas convention, the result is that the first half of a bit period matches the information bit and the second half is its complement. Power spectral density for Manchester is shown in fig. 3.

## 3 SYSTEM BLOCK DIAGRAM

The basic configuration of an SCM–WDM optical system is shown in fig. 4. In this example, independent high-speed digital signals are mixed by different microwave carrier frequencies. These are combined and optically modulated onto an optical carrier. Wavelengths are then multiplexed together in an optical WDM configuration. At the receiver, an optical demultiplexer separates the wavelengths for individual optical detectors. Then, RF coherent detection is used at the SCM level to separate the digital signal channels. Channel add–drop is also possible at both the wavelength and SCM levels. Although this SCM–WDM is, in fact, an ultra dense WDM system, sophisticated microwave and RF technology enables the channel spacing to be comparable to the spectral width of the baseband, which is otherwise not feasible by using optical technology. Compared to conventional high-speed TDM systems, SCM is less sensitive to fiber dispersion because the dispersion penalty is determined by the width of the baseband of each individual signal channel compared to conventional WDM systems, on the other hand, it has better optical spectral efficiency because much narrower channel spacing is allowed. Conventional SCM generally occupies a wide modulation bandwidth because of its double-sideband spectrum structure and, therefore, is susceptible to chromatic dispersion. In order to reduce dispersion penalty and increase optical bandwidth efficiency, optical SSB modulation is essential for long-haul SCM–WDM optical systems.

## 4 MATHEMATICAL ANALYSIS OF OBI

There are n numbers of subcarriers in a given optical channel as shown in fig. 1, having the same average power. Each of these fields can be represented by,

$$e_i = \sqrt{s_i(t)}$$

Where the intensity modulation by the RF subcarrier of center frequency $f_i$ is represented by

$$s_i(t) = 1 + m(t)\cos(2\pi f_i t)$$

Where m (t) represents NRZ, MC, and Manchester data stream with bit duration $\tau$ .

The total field in an optical channel is the sum of M field can be represented as

$$e(t) = \sum_{i=1}^{n} e_i(t)$$

The electric field at the output of the fiber is given by

$$e_\circ(t) = [e(t) \otimes h_f(t)]e^{-\alpha L}$$

Where $\alpha$ is the fiber attenuation coefficient, L is the fiber length and $h_f(t)$ represents the fiber impulse response. The photo detector converts this field into an electric signal proportional to the field intensity. Photo detector output current is then given by

$$i(t) = R|e_\circ(t)^2| + n(t) \quad (1)$$

Where R is responsivity of the photo detector (PD) and n(t) represent the noise due to photo detector and preamplifier. By expanding equation (1), signal component and cross component of photo detector output can be expressed as

$$i_s(t) = \sum_{i=1}^{n} e_i^2(t)$$

$$i_c(t) = \left\{ 2\sum_{i=1}^{n} \sum_{l=i+1}^{n} e_i(t)e_l(t) \right\}$$

Here $i_c(t)$ Contributes nonzero beat interference term. The output of the PD is passed through a preamplifier followed by a band pass filter. If any of the spectral component of $i_c(t)$ falls within the bandwidth of any one of the M users BPF, it will cause optical beat interference (OBI).

The frequency spectrum if the *i*-th subcarriers (channels) signal component can be obtained by taking Fourier transform of $i_{si}(t)$ as

$$I_{si}(f) = \int_{-\infty}^{+\infty} i_{si}(t)e^{-jft} df$$
$$= 2\pi\delta(f) + (\tau/2) * [\text{sinc}\{\tau(f+f_i)\} + \text{sinc}\{\tau(f-f_i)\}]e^{-jf\tau/2}$$

Where $f_i$ represents the *i*-it subcarrier frequency. The power spectrum of the *i*-th subcarrier's signal component can be expressed as

$$P_{si}(f) = [I_{si}(f)]^2$$
$$= 4\pi^2\delta(f) + (\tau^2/4)[\text{sinc}^2\{\tau(f-f_i)\} + \text{sinc}^2\{\tau(f+f_i)\}]$$

Using band pass filter, output signal power of the desired sub-carrier can be expressed as

$$P_{isig} = \int_{f_i-B/2}^{f_i+B/2} P_{si}(f) df$$
$$= \int_{f_i-B/2}^{f_i+B/2} (\tau^2/4)[\text{sinc}^2\{\tau(f-f_i)\}] df \quad (2)$$

Where B is the specified bandwidth of the sub-carrier or bandwidth of the BPF. Here [$\text{sinc}^2\{\tau(f+f_i)\}$] term falls outside range of integration. Similarly, the frequency spectrum of the composite cross component can be expressed as

$$I_c(f) = \int_{-\infty}^{+\infty} i_c(t)e^{-jft} df$$

The power spectrum of the composite cross component can be expressed as

$$P_c(f) = [I_c(f)]^2$$

The cross component power present in the required bandwidth B is

$$P_{icross} = \int_{f_i-B/2}^{f_i+B/2} P_c(f) df \quad (3)$$

This $P_{icross}$ is the source of the optical beat interference (OBI).

The signal to interference ratio (SIR) is defined as the ratio of the signal power in (2) to the power of the cross component in (3) within the required bandwidth. SIR can be expressed by the following equation

$$SIR = \frac{P_{isig}}{P_{icross}}$$

## 5  RESULTS

A theoretical performance analysis is presented for a subcarrier multiplexed (SCM) optical transmission system in presence of optical beat interference (OBI) which occurs during the photo detection process for non-return to zero (NRZ), delay modulation (DM) OR miller coding (MC) and Manchester linecoding. A multiuser SCM system for a suitable bandwidth is investigated using MATLAB and in this investigation allocated bandwidth for each user was 200 KHz with carrier separation 200 KHz. The results are shown below from fig. 5 to fig 17. The modulated RF carriers in SCM system then intensity modulate each optical carrier in WDM systems. Here we have used three types of line coding schemes NRZ, Manchester and Miller coding. Spectrum of signal and cross components respectively obtained using MATLAB. The signal and cross components were estimated using a 131072 point of FFT. Here we have plotted the comparative representation of SIR versus number of channels/subcarriers with NRZ, Manchester and Miller coding. We have found that the significant increase in the SIR for Miller coded SCM system compared to NRZ and Manchester for the same data rate with a given number of subcarrier channels and channel bandwidth. For the same optical modulation index with 2 multiplexed subcarrier channels each having a bandwidth of 200 KHz with channel spacing of 200 KHz, the available SIR is approximately -9 dB for NRZ coded SCM system and -14 dB for 2 multiplexed subcarrier channels for Manchester coded SCM, then the corresponding SIR available with MC coded SCM system is approximately -2 dB. For 10 multiplexed subcarrier channels the available SIR is -46 dB for NRZ coded system and -49 dB for Manchester coded system then the corresponding SIR is available -24 dB for MC system.

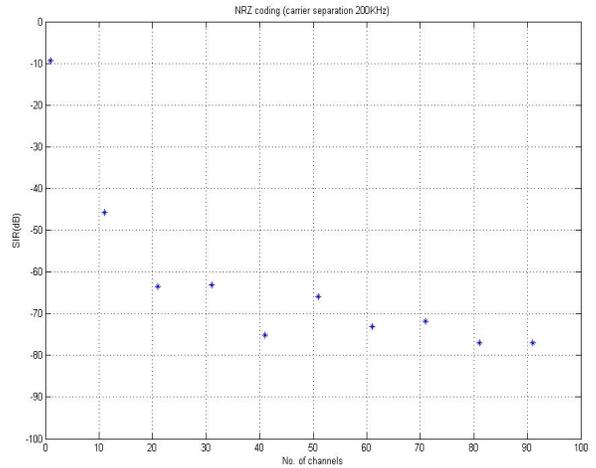

Fig. 7. SIR versus number of channels of an SCM system for NRZ.

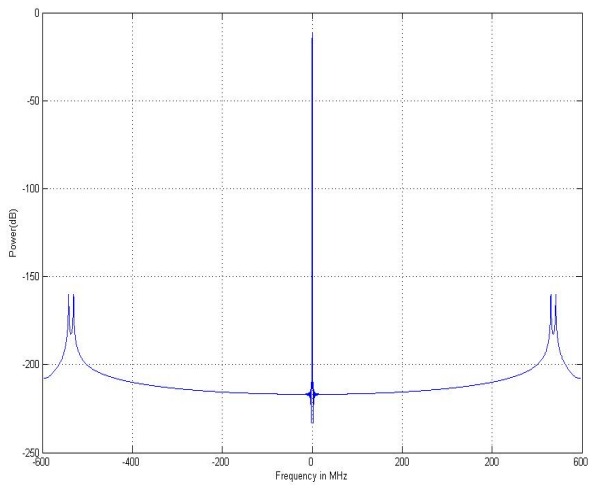

**Fig.** 5. Spectrum of signal component of an SCM system for NRZ.

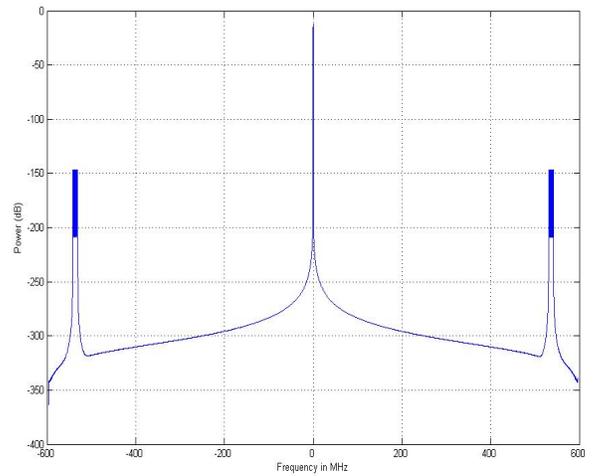

Fig. 8. Spectrum of signal component in an SCM system for Miller code.

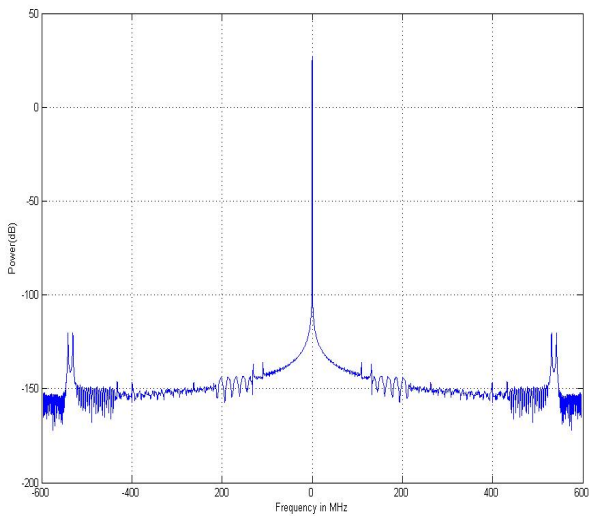

Fig. 6. Spectrum of cross component of an SCM system for NRZ.

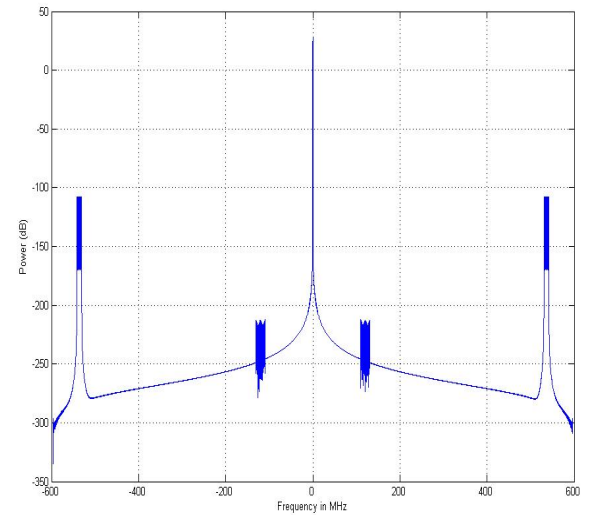

Fig. 9. Spectrum of cross component in an SCM system for Miller code.

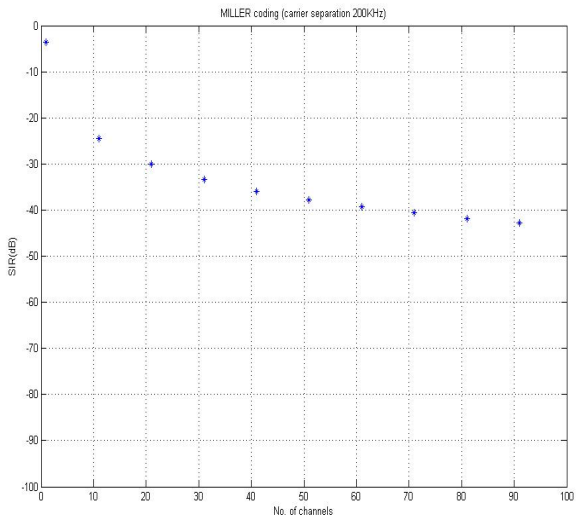

Fig. 10. SIR versus number of channels in an SCM system for Miller code.

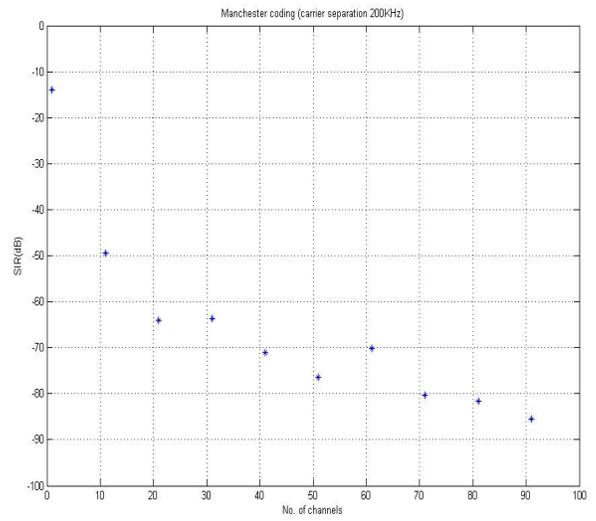

Fig. 13. SIR versus number of channels in an SCM system for Manchester.

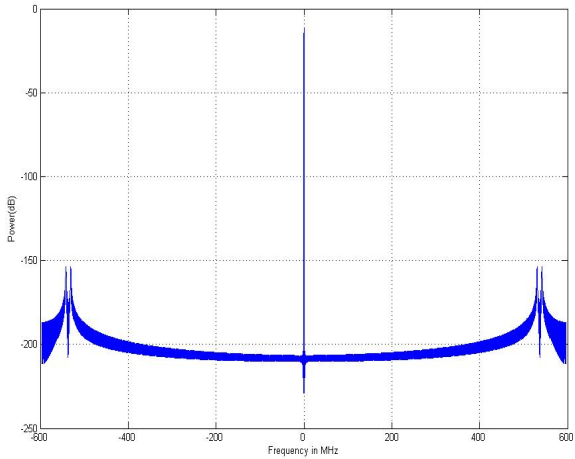

Fig. 11. Spectrum of signal component in an SCM system for Manchester.

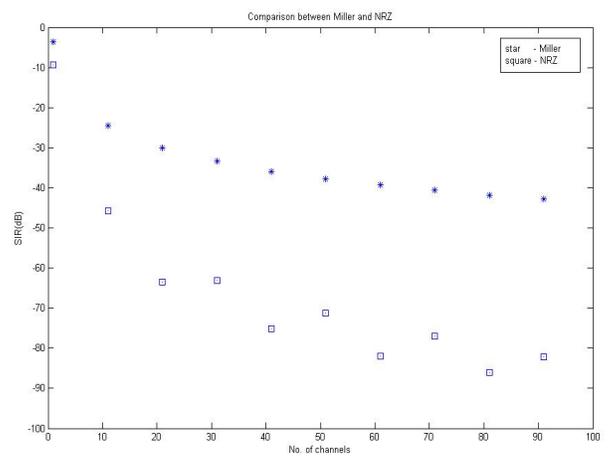

Fig. 14. SIR versus number of channels in an SCM system for Miller and NRZ.

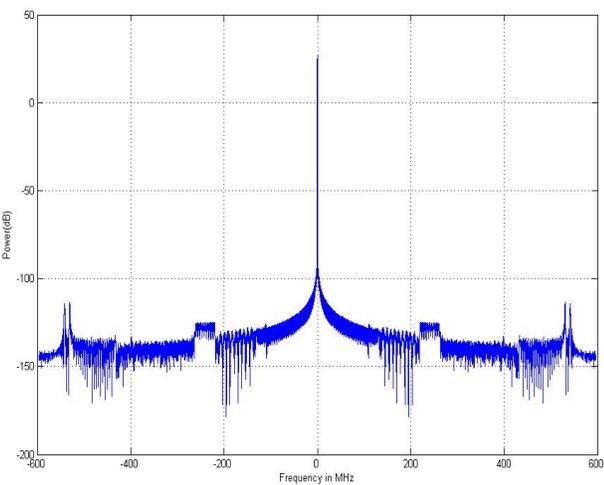

Fig. 12. Spectrum of cross component in an SCM system for Manchester.

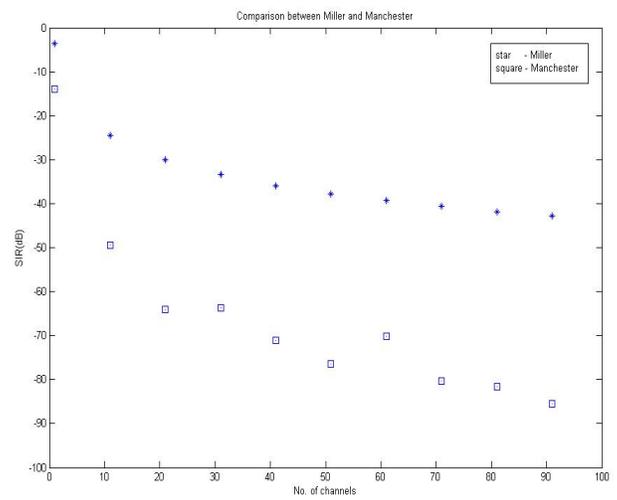

Fig. 15 SIR versus number of channels in an SCM system for Miller and Manchester.

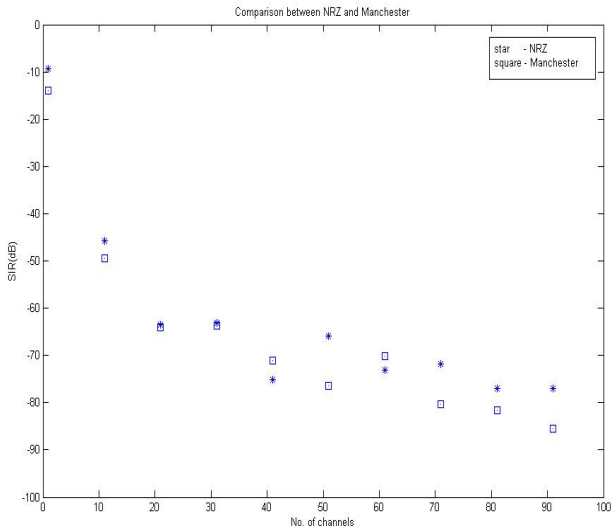

Fig. 16. SIR versus number of channels in an SCM system for NRZ and Manchester.

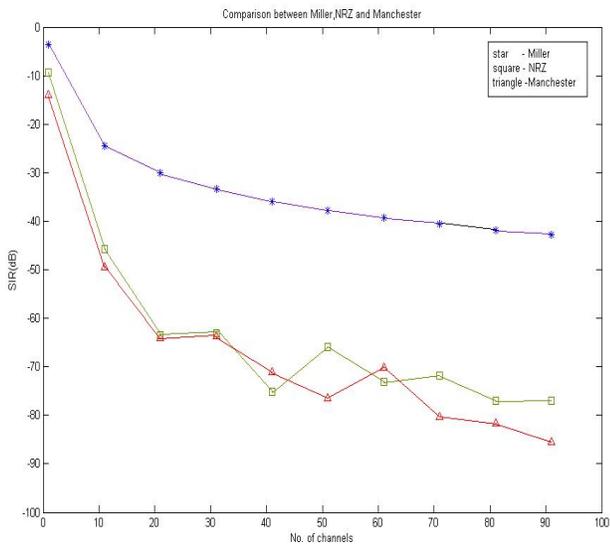

Fig. 17. SIR versus number of channels for Miller, NRZ and Manchester.

## 6   CONCLUSION

A theoretical analysis is presented to evaluate the limitations imposed by Optical Beat Interference on the performance of an SCM optical transmission system. Performance results reveal that OBI limits the number of subcarriers that can be multiplexed in an optical channel while employing NRZ, Manchester linecoding and Miller codes. However, by employing Miller codes, the limitations of OBI on SCM system performance can be reduced. The analysis easily shows that the performance of Miller code with respect to NRZ and Manchester linecoding is better. Here we have represented the comparative performance of three linecoding schemes in an SCM-WDM system.

There is more scope to analysis the performance of SCM system using another linecoding schemes such as Duobinary. Also it is possible to analysis the performance of an SCM system with respect to Bit Error Rate (BER).

**Md. Shamim Reza** was born in Magura, Bangladesh, on August 06, 1983. He received the B.Sc. and M.Sc. degrees in Electrical and Electronic Engineering from Bangladesh University of Engineering and Technology, Bangladesh, in 2006 and 2008, respectively. At present, he is an Assistant Professor with the Electrical and Electronic Engineering Department, Bangladesh University of Engineering and Technology, Dhaka, Bangladesh. His research interests include optical fibre communication, optical networking and signal processing.

**Md. Maruf Hossain** was born in Khulna, Bangladesh.. He received his B.Sc. Engg. degree in Electrical and Electronic Engineering from Bangladesh University of Engineering and Technology, Bangladesh, in 2006. Now he is doing his MSc Engg. in the Dept of Electrical and Electronic Engineering of Bangladesh University of Engineering and Technology, Bangladesh. At present, he is also a Lecturer with the Electrical and Electronic Engineering Department, American International University of Bangladesh, Dhaka, Bangladesh. His research interests include  optical fibre communication and wireless communications.

**Adnan Ahmed Chowdhury** is doing his BSc Engineering in the Dept of Electrical and Electronic Engineering, International Islamic University of Chittagong (IIUC), Dhaka Campus, Bangladesh. His Research interests include fibre communication and signal processing.

**S. M. Shamim Reza** is doing his BSc Engineering in the Dept of Electrical and Electronic Engineering, International Islamic University of Chittagong (IIUC), Dhaka Campus, Bangladesh. His Research interests include optical communication.

**Md. Moshiur Rahman** is doing his BSc Engineering in the Dept of Electrical and Electronic Engineering, International Islamic University of Chittagong (IIUC), Dhaka Campus, Bangladesh. His Research interests include fibre communication.